\documentstyle[sprocl]{article}

\input{psfig.sty}

\def\be{\begin{equation}}
\def\ee{\end{equation}}
\def\bea{\begin{eqnarray}}
\def\eea{\end{eqnarray}}
\def\tb{\tan\beta}
\newcommand{\mgaugino}{M_{1/2}}
\newcommand{\mt}{m_t}
\newcommand{\tg}{t_{\rm G}}
\newcommand{\mgut}{M_{\rm{GUT}}}

\newcommand{\lsim}{ \mathop{}_{\textstyle \sim}^{\textstyle <} }

\begin{document}

\title{NATURALNESS RE-EXAMINED:\\
IMPLICATIONS FOR SUPERSYMMETRY SEARCHES}

\author{J.~L.~FENG$^{a\dagger}$, 
        K.~T.~MATCHEV$^{b\ast}$ and 
        T.~MOROI$^{c\dagger}$}
\address{$^\dagger$School of Natural Sciences, Inst.\ for Advanced Study,
Princeton, NJ 08540, USA\\
$^\ast$Theoretical Physics Department, Fermilab,
Batavia, IL 60510, USA}
\address{E-mail: $^a$feng@ias.edu, $^b$matchev@fnal.gov,
$^c$moroi@ias.edu} 
\address{\rm{IASSNS-HEP-00/20, FERMILAB-Conf-00/062-T}}

\maketitle

\abstracts{ We discuss the origin of "focus points" in the scalar mass
RGEs of the MSSM and their implications for collider searches.  We
present a new exact analytic solution to the homogeneous system of
scalar mass RGEs in the MSSM for general $\tan\beta$.  This is then
used to prove that the focus point for $m^2_{H_u}$ depends only on the
value of the top Yukawa coupling at the {\em weak} scale (not its
value at the GUT scale) and is independent of the bottom Yukawa
coupling. }

\section{Focus Points in Supersymmetry}

Supersymmetry (SUSY) is the leading candidate for new physics beyond
the Standard Model. It provides a framework for naturally explaining
the stability of the weak scale with respect to radiative corrections.
In its minimal version, the Minimal Supersymmetric Standard Model
(MSSM), it also successfully predicts gauge coupling unification. Both
of these attractive features of SUSY hinge on the assumption that the
scale of the superpartner masses is not too far above the electroweak
scale. This has reinforced a widespread optimism that the next round
of collider experiments at the Tevatron, LHC or the NLC are guaranteed
to discover all superpartners, if they exists.

In SUSY,the weak scale is determined by the relevant model
parameters -- the soft SUSY breaking Higgs masses,
$m_{H_u}$ and $m_{H_d}$,
and the supersymmetric Higgs mass parameter, $\mu$:
\be
\frac{1}{2} m_Z^2 =
\frac{m_{H_d}^2-m_{H_u}^2 \tan^2\beta}{\tan^2\beta -1}
- \mu^2 \ \sim - m_{H_u}^2 -\mu^2,
\label{ewsb}
\ee
where the last relation holds for large $\tan \beta$.  
Naturalness requires that there be no
large cancellations in the RHS of eq.~(\ref{ewsb})
when expressed in terms of the {\em fundamental} parameters $\{a_i\}$
of the model (e.g., in minimal supergravity (mSUGRA),
$\{a_i\}=\{m_0, \mgaugino, \mu_0, A_0, B_0\}$).
Otherwise an unwanted hierarchy is reintroduced, and 
the appeal of SUSY as a solution to the 
hierarchy problem is lost.
The degree of fine-tuning involved in eq.~(\ref{ewsb})
is usually quantified in terms of the sensitivity
coefficients \cite{BG} 
$c_{a_i} \equiv \left| \partial \ln m_Z^2/ \partial\ln a_i \right|$.

\begin{figure}[t]
\begin{center}
\hspace{0.01in}
\psfig{figure=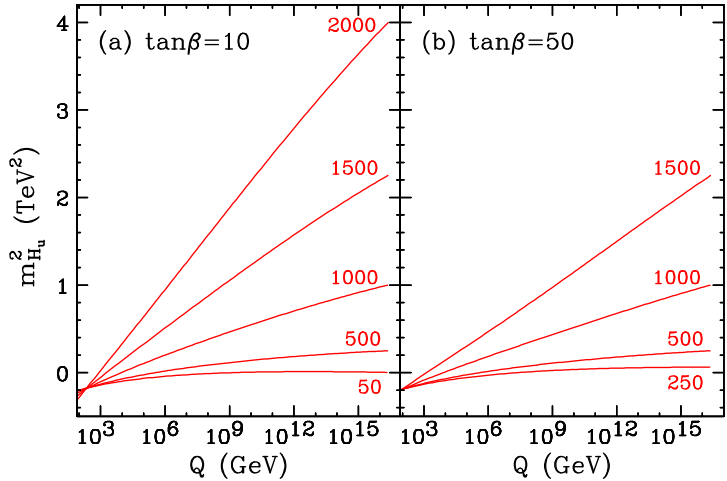,height=2.5in}
\end{center}
\caption{The RG evolution of $m_{H_u}^2$ for (a) $\tb=10$ and (b) $\tb=50$,
several values of $m_0$ (shown, in GeV), $\mgaugino = 300$ GeV,
$A_0=0$, and $\mt = 174$ GeV.  
\label{fig:run}}
\end{figure}

Recently we pointed out \cite{naturalness} that a general class of
supersymmetric theories, including mSUGRA, exhibits ``focus points''
in the MSSM renormalization group equations (RGEs). (See also
Ref.~\cite{Agashe}.) In particular, for the experimentally
interesting range of top quark mass values, $\mt\sim 170-180$ GeV,
$m_{H_u}$ has such a focus point at the weak scale and is therefore
highly insensitive to its GUT scale boundary value $m_0$ (see
Fig.~\ref{fig:run}).

\begin{figure}[t]
\begin{center}
\hspace{0.01in}
\psfig{figure=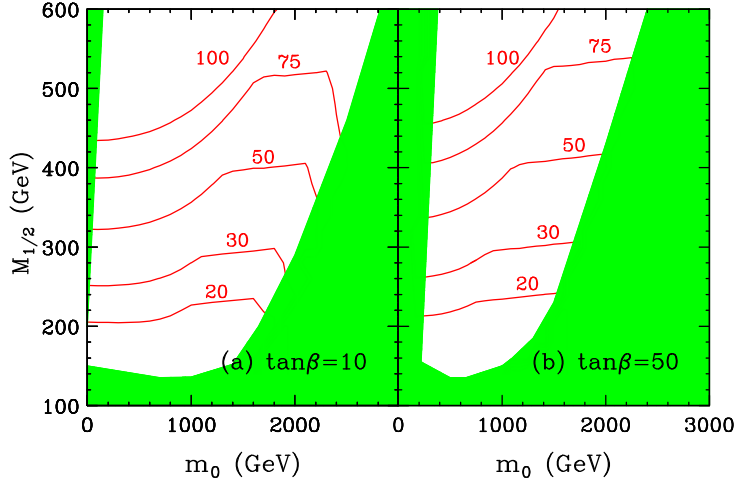,height=2.5in}
\end{center}
\caption{Contours of constant sensitivity parameter $c$ in the
$(m_0, \mgaugino)$ plane for (a) $\tb = 10$ and (b) $\tb = 50$,
$A_0=0$, $\mu>0$, and $\mt = 174$ GeV. The bottom and right shaded
region is excluded by the chargino mass limit of 90 GeV.  The top left
region is also excluded if a neutral LSP is required.}
\label{fig:m0M1/2}
\end{figure}

The focus point for $m^2_{H_u}$ implies that
$c_{m_0}$ is small \cite{naturalness}.
In Fig.~\ref{fig:m0M1/2} we show
contours of the overall sensitivity parameter
$c\equiv\max\{c_{a_i}\}$. We see that regions of parameter
space with $m_0\sim 2-3~{\rm TeV}$ are as natural as
regions with $m_0\lsim 1~{\rm TeV}$. Hence, it is
quite possible that all squarks and sleptons have
multi-TeV masses, making their discovery at colliders
significantly more challenging than conventionally expected.
On the other hand,
Fig.~\ref{fig:m0M1/2} reveals that naturalness restricts
the gaugino mass $M_{1/2}$ to the few hundred GeV range.

The focus point mechanism is surprisingly robust against variations of
the input parameters, e.g., $\mgaugino, A_0, \mt$ and $\mgut$.  We will
prove analytically\footnote{We are grateful to Jon Bagger for
stimulating questions and discussions.} that the focus point scale is
also independent of $\tan\beta$. This supplements previous numerical
demonstrations of this result~\cite{naturalness}.  In the process
(see section \ref{solution}) we shall derive analytic solutions to the
homogeneous parts of the scalar mass RGEs for arbitrary values of
$\tb$.  To our knowledge, such solutions have not been presented
previously in the literature.

\section{Analytic Solution to the Homogeneous System of Scalar RGEs}
\label{solution}

Neglecting the tau Yukawa coupling and
the hypercharge differences for top and bottom,
the RGEs for the relevant Yukawa couplings read
\bea
\dot{Y}_t = Y_t\ (6Y_t+Y_b-a_ig_i^2), \qquad
\dot{Y}_b = Y_b\ (6Y_b+Y_t-a_ig_i^2),  \label{dybdt} 
\eea
where $Y_t\equiv y_t^2$, $Y_b\equiv y_b^2$,
$8\pi^2t\equiv \ln Q$ and $\dot{}\equiv d/dt$. 
The homogeneous system of RGEs for the relevant scalar masses is
\be
\dot{\vec{M}}= {\bf N} \vec{M},
\label{scalar RGEs}
\ee
where $\vec{M}^T=(m^2_{H_u}, m^2_{U}, m^2_{Q}, m^2_{D}, m^2_{H_d})$
and 
\be
{\bf N} \ =\ Y_t
\left(
\begin{array}{ccccc}
3 &  3 &  3 &  0 &  0   \\
2 &  2 &  2 &  0 &  0   \\
1 &  1 &  1 &  0 &  0   \\
0 &  0 &  0 &  0 &  0   \\
0 &  0 &  0 &  0 &  0  
\end{array} \right) 
+ Y_b
\left(
\begin{array}{ccccc}
0 &  0 &  0 &  0 &  0   \\
0 &  0 &  0 &  0 &  0   \\
0 &  0 &  1 &  1 &  1   \\
0 &  0 &  2 &  2 &  2   \\
0 &  0 &  3 &  3 &  3 
\end{array} \right).
\label{scalar beta function}
\ee
Now perform a suitable change of variables 
\be
\left( \begin{array}{c}
m_{H_u}^2 \\
m_{U_3}^2 \\
m_{Q_3}^2 \\
m_{D_3}^2 \\
m_{H_d}^2
\end{array} \right)
=
\left( \begin{array}{rrrrr}
1 & 3 & 0 & 0 & 0 \\
-1& 2 & 1 & 0 & 0 \\
0 & 1 &-1 & 1 & 0 \\
0 & 0 & 1 & 2 & 1 \\
0 & 0 & 0 & 3 &-1
\end{array} \right)
\left( \begin{array}{c}
c_0 \\
c_t(t) \\
c_0' \\
c_b(t) \\
c_0''
\end{array} \right),
\label{change}
\ee
which amounts simply to decomposing $\vec{M}$ into
the eigenvectors of ${\bf N}$ (those are just the
columns of the $5\times5$ matrix in the RHS of
eq.~(\ref{change})). Since ${\bf N}$ has three zero eigenvalues,
three of the new variables, namely $c_0$, $c_0'$ and $c_0''$,
are constants and do not run. We have thus reduced the system
of five equations (\ref{scalar RGEs}) to a system of two equations
for the remaining new variables $c_t$ and $c_b$:
\bea
\dot{c}_t = Y_t\ (6c_t+c_b), \qquad
\dot{c}_b = Y_b\ (6c_b+c_t).  \label{dcdt}
\eea
The GUT scale boundary conditions (BC) are
\be
c_t(\tg) = k_t, \qquad
c_b(\tg) = k_b \label{cgut},
\ee
where $k_t$ and $k_b$ are some model-specific pure numbers.
In the absence of a general method for solving the
system (\ref{dcdt}), we make the following ansatz:
\begin{eqnarray}
c_t(t) = k_t+Y_t(t) f_t(t), \qquad
c_b(t) = k_b+Y_b(t) f_b(t),   \label{cansatz}
\end{eqnarray}
with $f_t(\tg)=f_b(\tg)=0$.  Then, we get
\begin{eqnarray}
\dot{f}_t&=& 6k_t+k_b+a_ig_i^2f_t-Y_b(f_t-f_b),  \label{dfdt}
\end{eqnarray}
and $\dot{f}_b$ is given by replacing the subscript $t\leftrightarrow
b$.  Now, by calculating $\dot{f}_t-\dot{f}_b$, we find a simple
equation for $f_t-f_b$,
\begin{equation}
(\dot{f}_t-\dot{f}_b)+(Y_t+Y_b-a_ig_i^2)(f_t-f_b) = 5(k_t-k_b),  
\label{dfmgdt}
\end{equation}
which can be easily integrated. To simplify notation,
let us define 
\be
Y(t) \equiv \int^t_{t_G} dt' (Y_t(t')+Y_b(t')),\qquad
G(t) \equiv \int^t_{t_G} dt' a_ig_i^2(t').       \label{YG}
\ee
Then the solution to eq.~(\ref{dfmgdt}) can be written as
\begin{equation}
f_t(t)-f_b(t)= 5(k_t-k_b)e^{G(t)-Y(t)}
\int^t_{t_G} dt' e^{-G(t')+Y(t')},  \label{fmg}
\end{equation}
where we have fixed the integration constant using $k_t=k_b$.  It is
now trivial to solve for $f_t(t)$ and $f_b(t)$ themselves.  The final
solution for $c_t(t)$ is
\begin{eqnarray}
c_t(t)&=& k_t+Y_t(t)e^{G(t)}\Biggl\{
(6k_t+k_b)\int^t_{t_G} dt' e^{-G(t')} \nonumber \\
    && - 5(k_t-k_b)
\int^t_{t_G} dt'  Y_b(t') e^{-Y(t')}
\int^{t'}_{t_G} dt'' e^{Y(t'')-G(t'')} \Biggr\}.  \label{ct} 
\end{eqnarray}
The solution for $c_b(t)$ is obtained simply by interchanging
$t\leftrightarrow b$ in (\ref{ct}).  As analytic solutions for $Y_t$
and $Y_b$ are known \cite{FL}, this then completes the analytic
solution of the homogeneous scalar mass RGEs {\em for arbitrary
$\tan\beta$}.

Now, we discuss the $\tan\beta$ independence of the focus point scale
of $m_{H_u}^2$.  Define the focus point scale $Q_{\rm F}^{(H_u)}$ as
\begin{eqnarray}
\left. \frac{dm_{H_u}^2}{dm_0} \right|_{Q=Q_{\rm F}^{(H_u)}} = 0.
\end{eqnarray}
The $m_0$ dependence of $m_{H_u}^2$ can be studied by using the
homogeneous part of the RGE for the squared scalar masses, and hence
$Q_{\rm F}^{(H_u)}$ is obtained by solving the following equation:
\begin{equation}
c_0 + 3 c_t(t_F) \equiv 0,
\label{focpoint}
\end{equation}
where $8\pi^2 t_F=\ln Q_{\rm F}^{(H_u)}$.  In mSUGRA we find
$c_0=-\frac{2}{7}m_0^2$, while $c_t(t_F)$ is given by eq.~(\ref{ct})
with $t=t_F$ and $k_t=k_b=\frac{3}{7}m_0^2$:
\begin{equation}
c_t(t_F) = \left( {3\over7}+3Y_t(t_F)e^{G(t_F)}
\int^{t_F}_{t_G} dt' e^{-G(t')} \right) m_0^2.
\label{ctfixed}
\end{equation}
The boundary condition $k_t=k_b$ leads to an extreme simplification,
as the second term in the brackets in eq.~(\ref{ct}) vanishes.
Substituting (\ref{ctfixed}) into eq.~(\ref{focpoint}), we obtain the
focus point condition in the form
\begin{equation}
Y_t(t_F)e^{G(t_F)} \int^{t_G}_{t_F} dt' e^{-G(t')}
= {1\over9}.
\label{fp2}
\end{equation}

The focus point may therefore be written in terms of only the gauge
couplings and $y_t(t_F)$.  The extremely mild $\tan\beta$ dependence
of $Q^{(H_u)}_{\rm F}$ for moderate to large $\tan\beta$ then follows
from $y_t(t_F)\propto 1/\sin\beta \sim {\rm const}+{\cal
O}(\tan^{-2}\beta)$.  Note that as $\tan\beta$ varies in this range,
$y_b$ and $y_t(t_G)$ will vary.  The variation in $y_t(t_G)$ was
claimed in Ref.~\cite{RS} to destroy the insensitivity of $m_{H_u}$ to
$m_0$.  We see, however, that this variation is irrelevant for the
focus point: the focus point depends on $y_t(t_F)$, not $y_t(t_G)$,
and for all moderate and large $\tan \beta$, the focus point remains
at the weak scale, and the sensitivity coefficient $c_{m_0}$ is small.

\section{Implications For Supersymmetry Searches}

We have seen that the naturalness bound on $m_0$ (i.e., the typical
sfermion mass) may be as large as a few TeV (see Fig.~\ref{fig:m0M1/2}).
This result has important implications for the superpartner spectrum
and, in particular, the discovery prospects for scalar superpartners
at future colliders. 

Multi-TeV sleptons (whose masses scale as $\sim m_0$), for example,
are beyond the kinematic limit $\sqrt{s}/2$ of all proposed linear
colliders and will also escape detection at hadron
colliders, as they are not strongly produced, and will
rarely be obtained in the cascade decays of strongly
interacting superparticles.
Multi-TeV squarks will, of course, also evade proposed linear
colliders.  
They will also stretch the LHC reach, although a
recent study \cite{squarks} in the lepton plus jets channel has
revealed discovery potential even for squarks up to 3 TeV.

In contrast to the sfermions, gauginos and higgsinos
cannot be very heavy in this scenario. For example, the
constraint $c\leq 50$ implies $M_{1/2}\lsim 400~{\rm GeV}$,
corresponding to $M_1\lsim 160~{\rm GeV}$, $M_2\lsim 320~{\rm
GeV}$, and $M_3\lsim 1.2~{\rm TeV}$.  Such gauginos will be
produced in large numbers at the LHC, and will be discovered in
typical scenarios. (However, the detectability of all gauginos
and higgsinos in cases of mass degeneracies at the LHC remains
an open question.) 

Finally, in spite of the relatively heavy squark masses allowed by the
focus point mechanism, the most natural range for the light Higgs mass
is somewhere below 118 GeV.  This is because large values of $|A_0|$
(and hence large stop mixing) are disallowed by naturalness, as they
induce too large $m_{H_u}^2$ through the RGEs.  Therefore, even
if all sleptons and squarks are very heavy, Run II of the Tevatron has
a golden opportunity to explore much of the most natural mSUGRA
parameter space in its search for the lightest Higgs boson.

\section*{Acknowledgments}
This work was supported in part by the Department of Energy
under contracts DE--FG02--90ER40542 and DE--AC02--76CH03000,
by the National Science Foundation under grant PHY--9513835,
through the generosity of Frank and Peggy Taplin (JLF),
and by a Marvin L.~Goldberger Membership (TM).

\end{document}